\documentclass[aps,a4paper]{revtex4}
\usepackage{graphicx}

\newcommand{\be}{\begin{equation}}
\newcommand{\ee}{\end{equation}}
\newcommand{\ba}{\begin{array}}
\newcommand{\ea}{\end{array}}
\newcommand{\bea}{\begin{eqnarray}}
\newcommand{\eea}{\end{eqnarray}}
\newcommand{\bdm}{\begin{displaymath}}
\newcommand{\edm}{\end{displaymath}}

\begin{document}

\title{Domains and Interfaces in Random Fields }

\author{Prabodh Shukla}%
\email{shukla@nehu.ac.in}
\affiliation{ Physics Department \\ North Eastern Hill
University \\Shillong-793 022, India.} 


\begin{abstract} 

We analyze the energetics of domains and interfaces in the presence of
quenched random fields, particularly at the lower critical dimension of 
the random field Ising model. The relevance of this study to experiments 
is also discussed.

\end{abstract} 

\maketitle

\large{

\section{Introduction} 

Uniformly ordered ferromagnetic states tend to break up into domains due
to a variety of reasons.  The most familiar reason is the competition
between short-range exchange forces, and long-range dipolar forces. The
domain structure has a lower free energy than the uniformly ordered
ferromagnetic state. Imry and Ma~\cite{imma} argued that quenched random
fields in a system may also cause a uniform ferromagnetic state to break
into domains. The main result of Imry and Ma is that there is a lower
critical dimensionality for a system below which an arbitrarily weak,
quenched random field with average value zero would destroy the uniform
ferromagnetic order even at zero temperature. They concluded that the
lower critical dimensionality is equal to two for spins of discrete
symmetry ( Ising spins ), and four for spins of continuous symmetry (XY or
Heisenberg spins). The argument of Imry and Ma is intuitively appealing,
but nonrigorous. It remained unclear for several years if the results
obtained from the argument were correct. The controversy was generated by
the existence of another argument based on a field theoretic method that
predicted the lower critical dimensionality of a pure ( i.e. not
disordered ) system to be two dimensions lower than the lower critical
dimensionality of corresponding random field system. The lower critical
dimensionality of pure Ising model is equal to one, and that of continuous
spin model is equal to two. Thus the dimensional reduction argument
predicted the lower critical dimensionality of random field Ising model to
be three, and that of random field continuous spin model to be four. It
took several years to realize the error in the dimensional reduction
argument, and to resolve that the results obtained on the basis of the
Imry-Ma argument are correct.

In section II, we define a classical spin model of ferromagnetism
incorporating random fields, and recapitulate the arguments due to Imry
and Ma for the existence of domains in the ground state of the model if
the dimensionality of the system is lower than a critical dimension.
Quenched random fields are actually realized in several experiments.  In
the following, we illustrate this with a few widely studied
examples~\cite{fishman,belanger}. We also mention briefly the difficulties
involved in the experiments, and in the interpretation of experiments. It
is due to these difficulties that the controversy between the domain wall
theory and the dimensional reduction theory could not be resolved for
several years based on experiments alone.

Although the domain argument gives the correct value of the lower critical
dimension, it becomes inconclusive exactly at the lower critical
dimension. It fails to clarify if Ising spins in two dimensions, and
continuous spins in four dimensions can support true long range
ferromagnetic order. In section III, we discuss in detail the random field
Ising model in two dimensions focusing on the nature of domain walls and
their roughness in this model. We conclude that there is no long range
ferromagnetic order in the two dimensional random field model. Our
presentation is largely based on the work of Binder ~\cite{binder}
supplemented with numerical simulation of the model. One might also wish
to examine more closely continuous spin models in four dimensions because
domain wall arguments are inconclusive here as well. However, we do not go
into this analysis, and not much is published on this problem. The reason
is perhaps historical. The conflict between the domain wall theory and the
dimensional reduction theory occurred in the case of lower critical
dimension of random Ising systems, and therefore the bulk of the study has
been on the random field Ising model.

\section{Domains in Quenched Random Fields }

We first describe the random field model for classical spins. A classical
spin is an n-component unit vector; it is called an Ising spin if n=1, XY
spin if n=2, and Heisenberg spin if n=3. Higher values of n may also
considered. The limit n $\rightarrow \infty$ is known as the spherical
model. Mutually interacting spins of continuous symmetry (n $\ge$ 2) do
not have a gap between the ground state and low lying excited states,
while the spins of discrete symmetry (n=1) do have a gap. Therefore the
qualitative behavior of continuous spin models is different from that of
the Ising model. For simplicity, we shall focus on the Ising model. The
model is defined on a d-dimensional hypercubic lattice. Each site is
labeled by an integer i, and carries a classical spin $S_{i}$ that takes
the value $+$1 if the spin is pointing along the positive z-axis, and $-$1
if it is pointing along the negative z-axis. Each site also carries a
quenched random magnetic field $h_{i}$. The set of fields $\{ h_{i} \}$
are independent identically distributed random variables with a continuous
probability distribution $\phi(h_{i})$. There is a ferromagnetic
interaction J between nearest neighbor spins. The Hamiltonian of the
system is

\be H=-J \sum_{i,j} S_{i} S_{j} - \sum_{i} h_{i} S_{i} \ee

In the absence of the random field, the ground state of the system at zero
temperature has all spins aligned parallel to each other. The ground state
is doubly degenerate. The energy of the system when all spins are pointing
up is the same as when they are all pointing down. To be specific, we
choose the state with all spins pointing up i.e. $S_{i}$ = 1 for each site
i, and focus on a block of linear dimension L inside the infinite system.
The volume of the block is $L^{d}$. The number of lattice sites and
therefore the number of spins contained in the block is also of the order
of $L^{d}$. The surface of the block is of the order of $L^{d-1}$. If the
spins in the entire block were to flip down simultaneously, the bulk
energy of the block would not be affected because the spins are still
aligned parallel to each other. However, the block would now appear as a
defect in the background of up spins. The energy of this defect will be
the surface energy of the block. The surface energy is of the order of
J$L^{d-1}$. In the absence of thermal energy or random fields, there is no
possibility that the block will flip down spontaneously in the fashion
envisaged above. Now consider the effect of random fields. The energy of
spin $S_{i}$ in the field $h_{i}$ is equal to $h_{i}S_{i}$ or simply
$h_{i}$ since $S_{i}$ = 1. Thus the random field energy of the block is
simply the sum of $L^{d}$ random fields inside the block. This sum is
equally likely to be positive or negative, and is of the order of $\sigma
L^{d/2}$, where $\sigma$ is the root mean square deviation of the random
fields from their average value zero. The surface energy of the block is
always positive. If the random field energy is negative and overwhelms the
surface energy, the block will flip down spontaneously to lower its
energy. Imry and Ma argue that if $ d/2 > (d-1) $, i.e. if $ d < 2 $, then
for any $\sigma$, there will be a characteristic length $L^{*}$ such that
the bulk energy gain will dominate over the surface energy loss for $L >
L^{*}$. In other words, domains will occur spontaneously if $ d < 2 $. The
characteristic length $L^{*}$ is given by the equation,

\be \left. {\sigma} {[L^{*}]}^{d/2} = J {[L^{*}]}^{d-1}, \mbox{ or } L^{*}
=\left[ \frac{J}{\sigma} \right] ^{\frac{2}{(2-d)}} \right. \ee

As may be expected, the characteristic size of the domain $L^{*}$
decreases with increasing disorder $\sigma$. Although we have considered
the formation of a single defect in a uniform ferromagnetic state,
numerous domains of linear size $L^{*}$ will be formed in the system if d
$<$ 2, and the uniform ferromagnetic state will be completely destroyed.
The density of domains is inversely related to their characteristic area.
On the basis of arguments presented above, Imry and Ma concluded that the
lower critical dimensionality of the random field Ising model is two, i.e.
the model would not support a uniform ferromagnetic state if the
dimensionality of the system is lower than two. The argument is
inconclusive in two dimensions.

Equation (1) is readily generalized to continuous spins. The Hamiltonian
is given by,

\be H=-J \sum_{i,j} \vec{S_{i}}. \vec{ S_{j}} - \sum_{i} \vec{h_{i}} .
\vec{S_{i}} \ee

The spin $\vec{S_{i}}$, and the field $\vec{h_{i}}$ are now n-component
vectors.  In the absence of the random field, the ground state of the
system at zero temperature is infinitely degenerate. We choose one of the
degenerate states, and consider the energetics of the formation of a
reversed spin block of size L. The random field energy of the block is
calculated as before, but the surface energy gets modified. In the case of
discrete spins the surface of the block is sharp. It cuts a pair of spins
that are oriented opposite to each other. Ising spins can only point up or
down. They cannot "tilt". The energy of a pair of Ising spins across the
wall is equal to J compared with $-$J in the absence of the wall.
Continuous spins can tilt and the system utilizes this flexibility to
lower its free energy. The change from up orientation outside the block to
down orientation inside the block is achieved gradually by bending away
from the z-axis over a distance that is known as the width of the domain
wall. The advantage of bending away from the z-axis is that each pair of
nearest neighbor spins in a domain wall is nearly parallel, and therefore
it is an energetically favorable way of creating a defect. Suppose the
angle between a pair of nearest neighbor spins in a domain wall is equal
to $\theta$. The energy of the pair with reference to the ground state
without any defect is therefore equal to $-2J(1- \cos{\theta})$, or ${J
{\theta}^2 }$ in the limit $\theta \rightarrow 0$. If the width of the
domain wall is $\omega$ ( i.e. there is a chain of $\omega$ bent spins
between an up spin and a down spin ), $\theta$ is equal to $\pi / \omega$,
and the energy cost of a wall of thickness $\omega$ is approximately equal
to $\omega$ $(\pi / \omega)^2$, or ${\pi}^2 / \omega$. This shows that the
energy cost of a wall goes to zero as the thickness of the wall goes to
infinity. In a real physical system, there is often an axis of easy
magnetization. The domains are oriented up or down along the axis of easy
magnetization. Bending of spins away from the easy axis costs what is
known as the anisotropy energy. The anisotropy energy increases with the
width of the wall, and therefore limits the width of the domain wall. In
our simple model, we have not considered the anisotropy energy term.
However, recognizing that domain walls have to be finite, we may make the
plausible assumption that the domain walls are of similar size as the
domains. In this case, the domain wall energy (per unit area of the wall)
is approximately equal to $J {\pi}^2 / L $.  Noting that the area of the
wall scales as $L^{d-1}$, we find that the domain wall energy scales as $J
L^{d-2}$. This is the energy cost of creating the domain. The gain from
random fields scales as $\sigma L^{d/2}$ as before. Comparing the two
terms we find the lower critical dimensionality for continuous spins is
equal to four. The characteristic length $L^{*}_{c}$ of domains in the
continuous spin model is given by the equation,

\be \left. {\sigma} {[L^{*}_{c}]}^{d/2} = J {[L^{*}_{c}]}^{d-2}, \mbox{ or
} L^{*}_{c} =\left[ \frac{J}{\sigma} \right] ^{\frac{2}{(4-d)}} \right.
\ee

\section{Relevance to Experiments}

A random field is not merely a theoretical artifact. Defects and
imperfections in magnets have the effect of creating a quenched random
field in the system. We illustrate this with a few examples
~\cite{fishman,belanger} that have played an important role in
investigating the effects of random fields experimentally.

The connection with experiments was first made by a theoretical argument
due to Fishman and Aharony ~\cite{fishman} who showed that a weakly
diluted anti-ferromagnet in a uniform external field acts like a
ferromagnet in a random field. Consider a one dimensional Ising
anti-ferromagnet. Each unit cell has two spins, say $S_{1}$ and $S_{2}$
that are oriented opposite to each other. The order parameter ( staggered
magnetization per unit cell ) is $S_{1} - S_{2}$, and the total
magnetization is $S_{1} + S_{2}$. The system is exposed to a uniform
external magnetic field $h$ that couples to the total magnetization. In a
randomly diluted system, one of the spins in a cell may be missing. Then
the external field would couple to the remaining spin instead of the sum.
The remaining spin may be written as an even or odd linear combination of
the total and staggered magnetization. We may write $S_{1} = [ (S_{1} +
S_{2}) + (S_{1} - S_{2}) ]/2$, and $S_{2} = [ (S_{1} + S_{2}) - (S_{1} -
S_{2}) ]/2$. This means that in cells where one spin is missing, the
external field couples to total magnetization as well as the staggered
magnetization. The sign of the coupling to the staggered magnetization (
the order parameter ) is positive if $S_{2}$ is missing, and negative if
$S_{1}$ is missing. The sign of the coupling between the order parameter
and its conjugate field is random because either spin can be absent with
equal probability. We may conclude from this that the critical behavior of
a randomly diluted anti-ferromagnet in one dimension will be similar to
that of a ferromagnet in a random field. The argument is generalizable to
higher dimensions. An anti-ferromagnet without dilution has two
sub-lattices on which the spins are oriented opposite to each other. One
of the sub-lattice is aligned with the applied field. In the presence of
dilution, locally the sub-lattice with most spins tends to align with the
applied field in competition with the global anti-ferromagnetic order in
the absence of dilution. As in the one dimensional case, the applied field
acts as an effective random field coupling to the order parameter (
staggered magnetization ).

The effective random field produced by the applied field is proportional
to the applied field. The strength of the random field is therefore easily
controlled. One can do scaling studies with varying strengths of disorder
in a system by simply tuning the applied field rather than making fresh
samples with different degrees of dilution. This simplifying feature has a
far reaching significance in investigating a theoretical model (RFIM)
experimentally, and the interaction between theory and experiment has
helped the field develop considerably, and clarified many fine points of
the model ~\cite{nattermann}.

If there is an experimental system that fits a theoretical model well, one
may think that the predictions of the model may be tested easily by
experiments. For example, the domain argument predicts the lower critical
dimensionality of RFIM to be two, and the dimensional reduction argument
predicts it to be three, so we may determine the correct value by doing an
appropriate experiment. However, the interpretation of experiments is
often not straightforward. Concentration gradients in the diluted sample
tend to round off a transition and affect the measurements of critical
behavior drastically. Further, the majority of experiments are performed
on samples prepared in the following two ways; (i) cooling it in zero
magnetic field, and (ii) cooling it in a magnetic field, and turning the
field off at the end. Experiments on samples prepared in the two ways give
different results. Three dimensional field cooled samples show no long
range order and were first thought to show that the lower critical
dimensionality of RFIM is three. But three dimensional samples cooled in
zero field showed long range order. So the question of lower critical
dimensionality could not be settled by experiments initially. It was only
after several years of controversy that the experimental situation
resolved itself in favor of the domain argument, i.e. no long range order
in two dimensions, but long range order in three dimensions. The main
point that was clarified by theoretical work is that the field cooled
state is not an equilibrium state. It relaxes logarithmically slowly, and
one should not expect to see an equilibrium ordered state in a field
cooled sample over any reasonable experimental time scale.

$Rb_{2}CoF_{4}$ is a good two dimensional Ising anti-ferromagnet
~\cite{ikeda}. It consists of layers of magnetic ions with a single
dominant intralayer exchange interaction, and an interlayer interaction
which is smaller by several orders of magnitude. It is very anisotropic so
that the spins can be well represented as Ising spins. The material can be
magnetically diluted by introducing a small fraction of manganese ions in
place of cobalt ions.  Crystals of $Rb_{2}Co_{x}Mn_{1-x}F_{4}$ are good
examples of a two dimensional diluted anti-ferromagnet suitable for an
experimental realization of the two dimensional random field Ising model.
In three dimensions, the most studied dilute anti-ferromagnet is
$Fe_{x}Zn_{1-x}F_{2}$ crystal. In the pure ferrous fluoride $FeF_{2}$
crystal ~\cite{hutchings}, the ferrous ions ( $Fe^{++}$ ) are situated
approximately on a body centered tetragonal lattice. Each ferrous ion is
surrounded by a distorted octahedron of flurine ( $F^{-}$ )  ions. The
predominant interactions are a large single-ion anisotropy, and an
anti-ferromagnetic exchange between nearest neighbor ferrous ions. The
magnetic moments of ferrous ions on the corners of the tetragonal cell are
anti-parallel to the magnetic moments of $Fe^{++}$ ions on the body
centers. The large crystal field anisotropy persists as the magnetic spins
are diluted with $Zn$. The diluted crystal remains an excellent Ising
anti-ferromagnet for all ranges of magnetic concentration $x$. Furthermore
crystals with excellent structural quality can be grown for all
concentrations $x$ with extremely small concentration variation $ \delta x
< 10^{-3}$. These attributes combine to make $Fe_{x}Zn_{1-x}F_{2}$ the
popular choice for experiments on diluted anti-ferromagnets, although
experiments have been done on several other materials as well
~\cite{belanger}.

\section{Roughness of Interfaces}

The discussion in section (2) was based on domains having flat interfaces
with other domains. For example, we considered a cube of upturned spins in
a uniform ferromagnetic background. The faces of the cube defined the
interface in this case. We could have considered a sphere of upturned
spins in which case the interface would have been a surface of constant
curvature. In both cases, the arguments for the stability of domains
presented earlier would hold good. The question arises if our conclusions
would hold even in the presence of rough interfaces with fluctuating
curvature. It is conceivable that the freedom of a wandering interface in
a random background would make it energetically more favorable for the
system to break up into domains. In other words, the roughness of the
interface could raise the lower critical dimensionality of the system from
the value given by the domain argument. In order to address this concern,
several workers reformulated the problem in terms of the interface. In
this approach we focus on a single interface in the system. It is
convenient to use a continuum model of the interface. We visualize the
interface as an elastic membrane. A flat interface is described by a
vector ${\vec{\rho}}$ in the (d-1)-dimensional hyperplane transverse to
the z-axis. Fluctuations in the interface are described by a displacement
$z(\rho)$ away from the flat surface ($z=0$). The displacement $z(\rho)$
is a scalar that can take positive or negative values. The random field is
denoted by a function $h(\rho,z)$ with the property,

\be < h(\rho_{1},z_{1}) h(\rho_{2},z_{2}) > =\delta(\rho_{1} -\rho_{2} )
\delta((z_{1}-z_{2}), \ee where $\delta$ denotes Dirac's delta function.

The Hamiltonian of the interface is written as,

\be H = J \int d{\vec{\rho}} [ 1 + ({\vec{\nabla_{\rho} }} z)^{2} ]^{1/2}
-\int d{\vec{\rho}} \int dz^\prime \mbox{ } h(\rho,z^\prime) \mbox{
sign}[z^\prime -z(\rho) ] \ee

The first term is the total area of the interface multiplied by the
exchange energy J; it gives the energy penalty for creating the interface.
The second term is the random field energy gained in roughening the
interface on the assumption that the spins are up above the interface and
down below it. The Hamiltonian can be simplified by adding to it a
constant term equal to

\bdm \int d{\vec{\rho}} \int dz^\prime \mbox{ } h(\rho,z^\prime) \mbox{ }
sign(z^\prime) \edm

One gets,

\be H = J \int d{\vec{\rho}} [ 1 + ({\vec{\nabla_{\rho} } } z)^{2} ]^{1/2}
+ 2 \int d{\vec{\rho}} \int_{0}^{z(\rho)} dz^\prime \mbox{ }
h(\rho,z^\prime)  \ee

On length scales larger than the bulk correlation length, the slope of the
interface $|\nabla_{\rho}|$ should be small in comparison with unity if
the domain argument is to be valid. In the following, we shall assume that
$|\nabla_{\rho}|$ is small, and explore the condition for the validity of
this assumption. We therefore linearize the fluctuations in the interface
to first order in $|\nabla_{\rho}|$. We also omit an infinite constant
that does not depend on the fluctuations in the interface. Thus we write

\bea H = \int d{\vec{\rho}} \left[ \frac{1}{2} J |{\vec{\nabla_{\rho} } }
z|^{2} + V(\rho,z(\rho)) \right] \cr \mbox{where, }V(\rho,z(\rho)) = 2
\int_{0}^{z(\rho)} dz^{\prime} h(\rho,z^{\prime}) \eea

Often the potential energy term $V(\rho,z(\rho))$ is regarded as an
independent random variable with mean value zero and variance given by,

\be < V(\rho_{1},z_{1}) V(\rho_{2},z_{2}) > =\delta(\rho_{1} -\rho_{2} )
R(z_{1}-z_{2}), \ee Here $R(z_{1}-z_{2})$ is a sort range function of its
argument ~\cite{grinstein,fisher}.

We characterize the roughness of the interface over a length $|\rho_{1}
-\rho_{2}|$ in the mean interface plane by a quantity $\omega(|\rho_{1} -
\rho_{2}|)$ that is defined as follows:

\be \omega^{2}(|\rho_{1} - \rho_{2}|) = \left< [z(\rho_{1})
-z(\rho_{2})]^{2} \right> \ee

Over large length scales, the roughness $\omega(|\rho_{1} - \rho_{2}|)$ is
characterized by an exponent $\zeta$ defined as follows:

\be \omega(|\rho_{1} - \rho_{2}|) = |\rho_{1} -\rho_{2}|^{ \zeta} \ee

If the exponent $\zeta$ is nonzero, but less than unity, the interface is
called a self-affine interface.

We shall not go into the detailed analysis of the interface Hamiltonian,
but be content to give rough scaling arguments to show that the interface
approach to the problem of lower critical dimensionality yields the same
result as the domain approach. We approximate
$|{\vec{\nabla_{\rho}}}z(\rho)|$ by $\omega(L)/L$, i.e. the ratio of the
height of the interface to its linear extension. In this rather crude
approximation the energy of the interface scales as

\bdm \Delta E = J L^{d-1} \left[ \frac{\omega}{L} \right]^{2} - \sigma
\sqrt{\omega L^{d-1} } \edm

The first term gives the energy cost of creating a fluctuation in a flat
interface and the second term the energy gain from the fluctuation (the
sum of random fields under the interface). The two terms should balance
each other in equilibrium. Differentiating $\Delta E$ with respect to the
height of the interface $\omega$, we get the scaling relations

\bdm \omega (L) = {\left[ \frac{\sigma}{4J} \right]}^{2/3}
L^{\frac{5-d}{3}}, \mbox{and } |{\vec{\nabla_{\rho}}}z(\rho)| =
\frac{\omega}{L} \sim L^{\frac{2-d}{3}} \edm

The above relations predict that the interface roughness is negligible if
d $>$ 2. The domain theory presented in section (2) assumes that the
interfaces are are not rough. We have now shown that this assumption is
justified if the dimensionality of the system is greater than two.
Actually, the interface approach is not fundamentally different from the
domain approach. Therefore the interface analysis presented above should
be seen as a consistency check with the domain approach, rather than an
independent proof of the results of the domain theory. The central result
of both approaches is that the lower critical dimensionality of the random
field Ising model is equal to two.

\section{Absence of Order in 2D RFIM}

Consider an L$\times$L square of down spins in a plane of up spins. Our
task is to investigate if this defect or a similar defect can appear
spontaneously in the two dimensional random field Ising model? The energy
of the defect (referred to the uniform state with all spins up) is equal
to

\bdm \Delta E = 8 J L - 2 \sigma L \edm

The first term is the energy required to create a domain wall along the
edges of the square. Each of the four edges of length L requires an energy
2 J L. The second term gives the possible gain in energy ( with
probability half ) from the random fields inside the square. There are
$L^{2}$ sites inside the square. The sum of random fields inside the
square is therefore approximately equal to $\pm$ $\sigma L$. The sum is
equally likely to be positive or negative. Only if the sum is negative,
the block gains energy by flipping down. However, if $\sigma < 4 J$, the
gain from random fields is not sufficient to compensate the loss from
domain walls. Therefore at first thought it appears that the long range
ferromagnetic order will remain stable in the presence of small disorder (
$\sigma < 4 J$ ).

We show in the following that sufficient energy may be gained from random
fields to compensate the loss due to domain walls if roughening of the
domain walls is taken into account. Suppose the square of down spins is
placed in the first quadrant of the xy plane such that the bottom left
corner of the square coincides with the origin of the coordinate system.
For simplicity, we focus on the roughening of just one edge of the square,
say the bottom edge along the x-axis from x=0 to x=L. Spins above the edge
are down, and those below it are up.

Following Binder ~\cite{binder}, we cut the wall into smaller pieces along
its length and allow each piece to move up or down without bending. The
cuts along the length are made in an iterative, self-similar manner on
smaller and smaller length scales. First the entire wall is allowed to
move up or down till it gains maximum energy from the random fields. Then,
the wall in its new position is cut into $n$ segments of equal length.
Each segment is allowed to move up or down in order to lower the energy of
the system. The process is repeated by cutting each segment into $n$ equal
pieces again, and relaxing each piece without allowing it to bend.

Suppose the entire length L of the horizontal wall moves vertically by a
distance $\omega$ when relaxed. The area covered by the moving wall is
equal to $\omega L$, and therefore the gain from random fields is equal to
$- \sigma \sqrt{ \omega L} $. The wall may move up or down with equal
probability. We consider the case when the wall moves down. In this case
the patch of down spins increases in area from an L$\times$L square to a
rectangle L$\times$(L+$\omega$) rectangle, and an extra vertical wall of
length $\omega$ has to be created on each side of the moving wall.  This
costs an energy equal to 2 J $\omega$ on each side. The total energy
change in relaxing the wall is equal to,

\be \Delta E = 4 J \omega - \sigma \sqrt{ \omega L} \ee

Minimizing the above expression with respect to $\omega$ gives

\be \omega = {\left( \frac{\sigma}{8J} \right)}^{2} L \ee

Substituting from equation (13) into equation (12) gives,

\be \Delta E = - 4 J{\left( \frac{\sigma}{8J} \right)}^{2} L \ee

Next we divide the wall into $n$ equal segments. Each segment of length
$L/n$ is relaxed independently. It moves a distance $\omega_{1}$, and
gains an energy for the system equal to $\Delta E_{1}$ given by,

\bea \omega_{1} = {\left( \frac{\sigma}{8J} \right)}^{2} \frac{L}{n};
\mbox{ } \Delta E_{1} = - 4 J {\left( \frac{\sigma}{8J} \right)}^{2}
\frac{L}{n} \eea

Dividing each segment of length $L/n$ into further $n$ equal segments, and
relaxing each smaller segment independently yields,

\bea \omega_{2} = {\left( \frac{\sigma}{8J} \right)}^{2} \frac{L}{n^{2}};
\mbox{ } \Delta E_{2} = - 4 J {\left( \frac{\sigma}{8J} \right)}^{2}
\frac{L}{n^{2}} \eea

The process is continued till the typical distance moved by the smallest
segment is of the order of unity. Note that distances less than unity ( in
units of lattice constant ) do not have any physical significance on a
lattice. Suppose it takes kmax iterations to reach the end, i.e.

\be \omega_{kmax} = {\left( \frac{\sigma}{8J} \right)}^{2}
\frac{L}{n^{kmax}} = 1, \ee or,

\be kmax= \left[ \frac{2 \log \left( \frac{\sigma}{8J} \right) + \log L}
{\log n} \right] \ee

The energy gained in the entire process of relaxation is given by,

\be \Delta E = \sum_{k=0}^{k=kmax} n^{k} \Delta E_{k} = - 4 J {\left(
\frac{\sigma}{8J} \right)}^{2} L \times (kmax+1) \ee

Or,

\be \Delta E = - 4 J {\left( \frac{\sigma}{8J} \right)}^{2} L \times
\left[ \frac{2 \log \left( \frac{\sigma}{8J} \right) + \log L + \log n}
{\log n} \right] \ee

In the limit $\sigma \rightarrow 0$, and $L \rightarrow \infty$, the
dominant term becomes

\be \Delta E = - 4 J {\left( \frac{\sigma}{8J} \right)}^{2} \left[ \frac{L
\times \log L} {\log n} \right] \ee

Thus the energy gained from random fields by roughening the domain scales
as $L \log L$. Roughening of the flat domain wall also increases its
length. However, the increase in the length of the domain wall due to
roughening is marginal in cases where the domain wall is well defined. In
order to see this, we estimate the new length of the staggered contour
describing the interface:

\bea L_{new} = L + \sum_{k=0}^{k=kmax} n^{k} |\omega_{k}| = L [1 + {\left(
\frac{\sigma}{8J} \right)}^{2} (kmax+1)] \\ = L \left[ 1 + {\left(
\frac{\sigma}{8J} \right)}^{2} . \left\{ \frac{2 \log \left(
\frac{\sigma}{8J} \right) + \log L + \log n} {\log n} \right\} \right]
\eea

The first term is the leading term in the limit $L \rightarrow \infty$,
$\sigma \rightarrow 0$. The maximum excursion of the rough interface away
from its initial flat position is estimated as,

\be \omega_{max} = \sum_{k=0}^{k=kmax} |\omega_{k}| = {\left(
\frac{\sigma}{8J} \right)}^{2} L \sum_{k=0}^{k=kmax} n^{-k} \approx
{\left( \frac{\sigma}{8J} \right)}^{2} L \ee

The average square deviation from the flat interface is given by,

\be < \omega^{2}(L)  > = \sum_{k=0}^{k=kmax} {\omega_{k}}^{2} = {\left(
\frac{\sigma}{8J} \right)}^{4} L^{2} \sum_{k=0}^{k=kmax} n^{-2k} \approx
{\left( \frac{\sigma}{8J} \right)}^{4} L^{2} \ee

The characteristic length scale $L^{*}$ over which the energy gained by
the rough interface becomes comparable to the cost of making the interface
is given by the equation,

\bdm \Delta E \approx 2 J L - 4 J {\left( \frac{\sigma}{8J} \right)}^{2}
\left[ \frac{L \times \log L}{\log n}\right] \approx 0 \edm

The coefficients of the two terms in the middle are not intended to be
exact. The important point is that the first term ( the cost of making a
domain wall ) is linear in L, and the second term ( the grain from
roughening the wall in random fields ) is of the order of $L \log L$. The
characteristic length $L^{*}$ is the minimum length beyond which $\Delta
E$ can become negative, and therefore the system can spontaneously break
up into domains. Approximately,

\bdm L^{*} \approx {\left( \frac{\sigma}{J} \right)}^{2} L^{*} \times \log
L^{*}; \mbox{ or } \log L^{*} \approx \left( \frac{J}{\sigma} \right)^{2};
\mbox{ or } L^{*} \approx \exp{ \left( \frac{J}{\sigma} \right)^{2} } \edm

It is easily checked that over length scales $L^{*}$, the total length of
the staggered contour of the interface is of the order $L^{*}$ because ${(
\sigma / J )}^{2} \log L^{*}$ is of the order unity. This proves an
assumption made earlier that there is a length scale over which the cost
of the domain wall scales as L, and the gain from random fields as $L \log
L$. The gain from random fields dominates over the cost of making a domain
wall for $L > L^{*}$, and the system forms domains spontaneously. Hence we
conclude that the two dimensional random field Ising model does not
support long range ferromagnetic order in the limit of infinite system
size.

A few remarks are in order:
\begin{enumerate}

\item The energy gained from random fields depends on a parameter $n$ that
is an artifact of our method of estimation of the energy. However this
dependence is rather weak because it arises via a $\log n$ factor. It is
not expected to seriously affect the conclusions drawn above.

\item Excursion of the wall at the k-th step may take it into a region
already crossed by the wall at the previous step. Thus $\Delta E_{k}$ may
have a contribution that is already included in $\Delta E_{k-1}$. In other
words, there may be a problem of double counting in our procedure. Binder
~\cite{binder} argues that this problem is not serious.

\item Binder ~\cite{binder} also argues that the conclusions drawn above
would hold even if all cuts of the flat interface at the k-th stage are
not exactly equal to $L / n^{k}$, but are equal to this value on the
average.

\end{enumerate}

\section{A Toy Problem}

The analysis presented in the preceding section shows that the energy
gained by an interface from roughening in a random background scales as L
log L, where L is the length of the interface. This result is extracted
after several simplifying assumptions and approximations. It is therefore
desirable to check the theoretical prediction against numerical
simulations of the model. The difficulty is that the numerical simulation
of the minimum energy configuration of an interface in a random background
is almost as difficult as an exact theoretical solution of the problem. We
therefore adopt an algorithm that has the virtue of being efficient, and
we hope that it captures the essence of the physical problem. Simulation
of our toy problem on systems of moderate size appear to bear out the
theoretical predictions of the previous section ~\cite{deepak}.

We consider a square lattice of size L$\times$L, and an interface that
joins the bottom left corner of the square to the top right corner. It is
convenient to consider the bottom left corner as the origin of the
coordinate system; the bottom edge of the square as the x-axis; and the
left edge as the y-axis. The interface is modeled by the path of a
directed walk along the nearest neighbor bonds on the lattice. The walk is
always directed along the positive y-axis and the positive x-axis. The
interface made by such a walk does not have any overhangs or loops. For
simplicity, we assume that the spins above the interface are down, and
those below it are up. The spins reside in the middle of plaquettes made
by joining nearest neighbor bonds around square unit cells. The spins are
conveniently labeled by cartesian coordinates $(i,j)$ where i denotes the
i-th row from below, and j-th column from left. A spin in the plaquette
$(i,j)$ experiences a random field $h(i,j)$.

Our object is to find an interface that divides the up spins from the down
spins in such a way that the net interaction energy of the spins with the
random fields at their sites is minimum. We consider the spins in one
column of the square at a time. Suppose the interface in the first column
separates $i$ spins below the interface that are up, from $L-i$ spins
above the interface that are down ($0 \le i \le L$). The magnetic energy
of the first column can be written as,

\be E(i,1) = \sum_{i^\prime=0}^{i^\prime=L} h(i^\prime,1) S_{i^\prime,1} =
\sum_{i^\prime=0}^{i^\prime=i} h(i^\prime,1)  -
\sum_{i^\prime=i+1}^{i^\prime=L} h(i^\prime,1) \ee

A computer algorithm generates random fields $h(i,1)$; and computes
$E(i,1)$ for each row in the first column ($i=1,2, \ldots L; j=1$). The
same process is repeated for the second column, i.e. we generate random
fields $h(i,2)$, and compute $E(i,2)$ for each row in the second column.
Now a naive approach may be to put the interface at a position in the
first column that corresponds to the minimum energy in the set
$\{E(i,1)\}$, and put the interface in the second column that corresponds
to the minimum energy in the set $\{E(i,2)\}$ subject to the restriction
that the interface in the second column be placed at the same level or
above the interface in the first column. Our algorithm described below is
an improvement over this naive approach. We compute a quantity
$E_{min}(i,2)$ for each row in the second column from $E(i,2)$ and
$\{E(i,1)\}$ as follows:

\be E_{min}(i,2) = Min (0 \le j \le i) \left[ E(i,2) + E(j,1) \right] \ee

This process is recursed to calculate $E_{min}(i,3)$ from $E(i,3)$ and
$\{E_{min}(i,2)\}$; and so on.

\be E_{min}(i,3) = Min (0 \le j \le i) \left[ E(i,3) + E_{min}(j,2)
\right] \ee

After L recursions of the process we get,

\be E_{min}(i,L) = Min (0 \le j \le i) \left[ E(i,L) + E_{min}(j,L-1)
\right] \ee

The quantity $ E_{min}(L,L) $ gives the minimum energy required to place
an interface diagonally across the square. Equivalently, $ E_{min}(L,L) $
is the maximum energy that an interface whose length is of the order of L
can gain by roughening itself in the background of quenched random fields.

Figure 1 shows the results of numerical simulations on systems of linear
size varying from $L=5$ to $L=100$. $E_{min}(L,L)/L$ is plotted along the
y-axis against $\log L$ on the x-axis. The figure shows graphs for three
distinct distributions of the random field: (i) random fields taking two
discrete values $\pm1$ with equal probability, (ii) a continuous
rectangular distribution of width 2 centered around the origin, and (iii)  
a gaussian distribution of variance unity. The data has been averaged over
$10^6$ independent realizations of the random field distribution. The
figure suggests that the energy gained by the interface from random fields
scales as $L \log L$. The data for the rectangular distribution has been
multiplied by $\sqrt{3}$. The modified data for the rectangular
distribution (after multiplication by $\sqrt{3}$) lies very close to the
data for the discrete $\pm 1$ distribution. This may be expected because
the standard square deviations of the two distributions are in the ratio
$1:\frac{1}{3}$.

\section{Thermal Effects}

At finite temperatures, interfaces are roughened by thermal energy as well
as by quenched disorder. Let $\xi_{d}$ be the length scale that describes
roughening due to disorder at zero temperature, and $\xi_{T}$ be the
characteristic length over which the interface would roughen due to
temperature if there were no disorder. As discussed in the previous
section, $\xi_{d}$ is the smallest portion of a straight line interface
that moves sideways by a unit distance when the interface is relaxed in a
random background. We get from equation (17), 

\be \xi_{d} = {\left ( \frac{8J}{\sigma} \right ) }^{2} \ee

The thermal length $\xi_{T}$ is the typical distance one has to move along
a straight interface before one finds a kink (i.e. a sideways turn of the
interface) due to thermal fluctuation. The density of kinks is
proportional to the Boltzmann factor $\exp{(-2J/k_BT)}$. Therefore,

\be \xi_{T} = \exp \left ( \frac{2J}{k_BT} \right ) \ee

Let us consider a system with fixed disorder $\sigma$, and raise its
temperature gradually above zero. At very low temperatures we have
$\xi_{T} >> \xi_{d}$ and therefore roughening of the interface is
dominated by disorder. As may be expected, thermal fluctuations are
essentially irrelevant at very low temperatures. At higher temperatures,
when $\xi_{T} < \xi_{d}$, the thermal effects become relatively important
and enhance roughening of the interface. Consider roughening by disorder
at length scale $L_{k}$. The straight segment of length $L_{k}$ will
develop on average $L_{k}/ \xi_{T}$ kinks due to thermal fluctuations.
Thus the mean square deviation $<\omega^{2}>_{T}$ due to thermal effects
will be equal to

\be <\omega^{2}>_{T} = \frac{L_k}{ \xi_{T}} = L_{k} \exp -
\left(\frac{2J}{k_BT} \right) \ee

The probability distribution of the interface width may be written as

\be P_T(\omega) \approx \exp \left[ - \frac{\omega^2}{2<\omega^{2}>_{T}} +
\frac{\sigma \sqrt{\omega L_k}}{k_BT} \right] \ee

The first term on the right is the distribution corresponding to a
completely random (Gaussian) walk. The second term takes into account the
energy that can be gained by excursions in a random field background.
Minimizing $P_T(\omega)$ with respect to $\omega$ gives

\be \omega_l = \left( \frac{\sigma}{T} \right)^{\frac{2}{3}} L_{k} \exp
\left( - \frac{4J}{k_BT} \right) \ee

The energy gained from random fields at length scale $L_k$ is given by

\be U_k = -\sigma \sqrt{\omega L_k} = - \sigma \left( \frac{\sigma}{T}
\right)^{\frac{1}{3}} L_{k} \exp \left( - \frac{2J}{3k_BT} \right) \ee

Summing over contributions from all independent length scales, we get the
total gain in energy U from roughening of a domain wall in a random medium
at a finite temperature

\be U = -\sigma \sqrt{\omega L_k} = - \sigma \left( \frac{\sigma}{T}
\right)^{\frac{1}{3}} L \log L \exp \left( - \frac{2J}{3k_BT} \right) \ee

The leading term in the cost of making a domain wall still scales linearly
with the length of the wall. At a characteristic length $L^{*}$ the energy
cost and energy gain terms balance each other. We get

\be L^{*} = \exp \left[ \left( \frac{J}{\sigma} \right) \left(
\frac{T}{\sigma} \right)^{\frac{1}{3}} \exp \left( \frac{2J}{3k_BT}
\right) \right] \ee

As noted earlier in this section, the above equation applies in the regime
$ \exp (J/k_BT) < (8J/ \sigma)$. Random field systems with linear size
greater than $L^{*}$ will spontaneously breakup into domains at
temperature T. Our discussion assumes that we have an interface with no
overhangs and bubbles. This assumption is justified if the density of
kinks is negligible, or equivalently if $\xi_{T} >> 1$. If overhangs and
bubbles are present in the system, it is not possible to formulate the
problem in terms of a single-valued variable $z(\rho)$ describing the
transverse fluctuations of a wall from a (d-1)-dimensional flat reference
plane.

At higher temperatures, at around the critical temperature of the pure
(i.e. without disorder)  system overhangs and bubbles are no longer
negligible. At these temperatures, a new length $\xi_c$ comes into play
that describes the diverging correlation length of the pure system at its
critical point. We have,

\be \xi(t)=t^{-\nu}, t=\frac{T_c-T}{T_c}, \ee where $T_c$ is the critical
temperature of the pure system, $t$ is the reduced temperature, and $\nu$
is a critical exponent. In a domain of linear size $\xi_c$, the net random
field is of the order of $\sigma \xi^{d/2}$, but the spins within the
domain are not aligned parallel to each other to the same degree as at low
temperatures. The average magnetization per spin is of the order of
$t^\beta$, where $\beta$ is another critical exponent. Thus the gain from
random field energy is reduced to a quantity of the order of

\be U = \sigma \xi^{d/2} t^{\beta} = \sigma \xi^{\frac{d \nu -2 \beta}{2
\nu}} = \sigma \xi^{\frac{\gamma}{2 \nu}} \ee

Here $\gamma$ is the critical exponent describing the diverging
susceptibility at the critical temperature. We have made use of the well
known relations, $\alpha + 2 \beta + \gamma =2$, and $d \nu = 2 - \alpha$,
where $\alpha$ is the critical exponent describing the singular part of
the specific heat. Just as the thermal fluctuations reduce the gain from
random fields by a factor $t^{\beta}$, they also reduce rather drastically
the cost of making a domain wall. The domain walls become ill-defined and
spread out over a distance comparable to the linear size of the domain,
and therefore the cost of making a wall is reduced from $J \xi$ to $J$
only. The cost and gain become comparable to each other at a value of
$\xi$ given by

\be \xi = \left( \frac{J}{\sigma} \right)^{\frac{2 \nu}{\gamma}} \ee

The crossover to the length scale $\xi$ takes place at a temperature given
by the reduced temperature

\be t =  \left( \frac{\sigma}{J} \right)^{\frac{2}{\gamma}} \ee

In summary, as the temperature of the system is increased keeping the
amount of random disorder $\sigma$ fixed, we may see crossover phenomena
between different length scales; a crossover from $\xi_d$ to $L^{*}$,
another from $L^{*}$ to $\xi$, and finally from $\xi$ to a distance of the
order of unity at very high temperatures. It is plausible that the length
scales are rather asymmetrical on the two sides of the transition
temperature $T_c$ of the pure system, and they remain finite and smaller
than $L^{*}$ in the vicinity of $T_c$ due to slow relaxation of the system
on experimental time scales. This speculation may possibly explain several
observed phenomena in systems where the underlying lattice goes through a
transition driven by a competition between temperature and quenched
disorder.

I thank D Dhar for critically reading the manuscript. 

}

\begin{figure}

\begin{center}

\includegraphics[angle=-90,width=16cm]{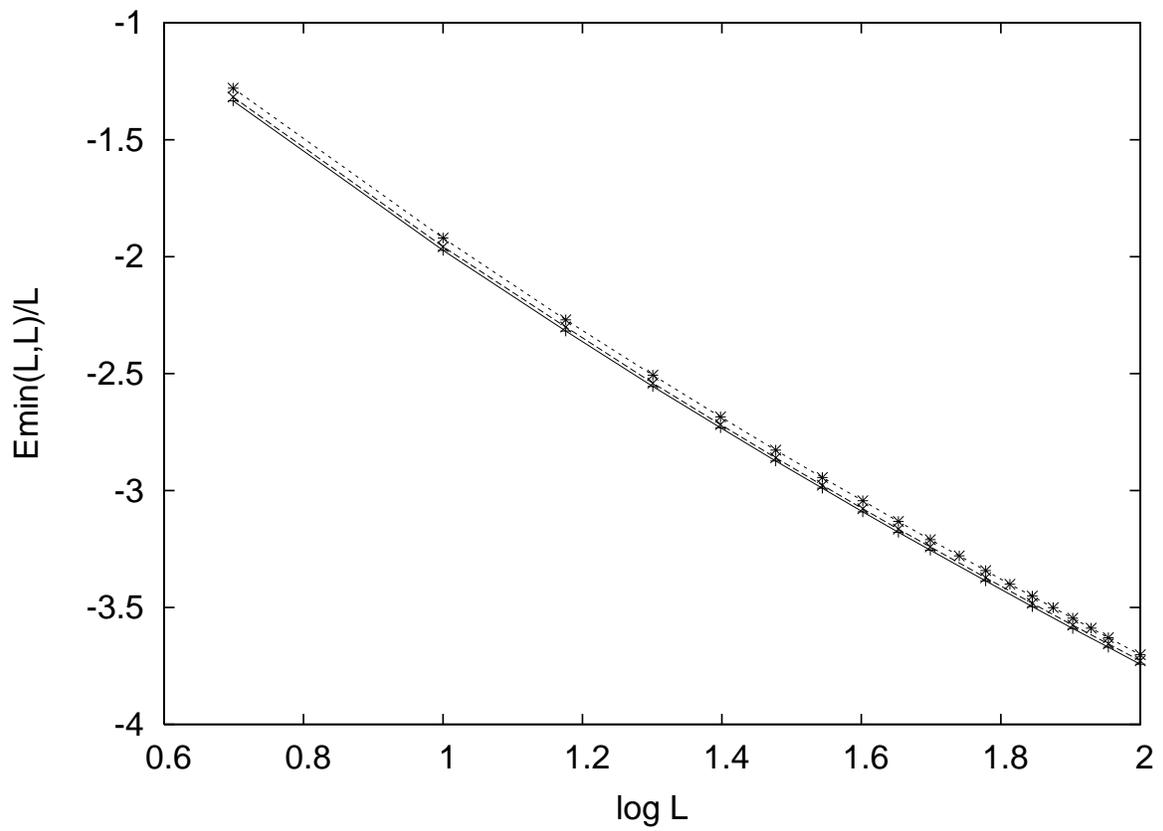}

\caption{ Energy gained by roughening of a diagonal interface in an
$L\times L$ square. The energy gain divided by L is plotted along the
y-axis against $\log L$ on the x-axis. Simulations for three different
random field distributions are shown. The lower curve is for a discrete
distribution $h_i=\pm J$; the middle curve for a rectangular distribution
of width $2J$ centered around the origin;  the upper graph is for a
Gaussian distribution of variance J. We have set $J=1$, and scaled the
middle curve by a factor $\sqrt{3}$ (see text).The figure suggests that
the energy gain scales as $L \log L$. }

\end{center}

\end{figure}


\begin{thebibliography}{99}
\bibitem{deepak} The idea of simulation was suggested by Prof D Dhar and
carried out by Sumedha and the author during the course of these lectures.
\end{thebibliography}
\end{document}